\documentclass{article}
\usepackage{spconf,amsmath,graphicx}
\usepackage{multirow}
\pdfoutput=1
\usepackage{hyperref}
\hypersetup{
    colorlinks=true,
    linkcolor=blue,
    filecolor=magenta,      
    urlcolor=cyan,
    pdftitle={A Multi-task convolutional neural network for blind stereoscopic image quality assessment using naturalness analysis},
    pdfpagemode=FullScreen,
    }

\title{A Multi-task convolutional neural network for blind stereoscopic image quality assessment using naturalness analysis}
\name{Salima Bourbia$^{1}$, Ayoub Karine$^2$, Aladine Chetouani$^3$, Mohammed El Hassouni$^{1,4}$}

\begin{document}
%
\maketitle
\begin{abstract}
\vspace{-2mm}
\ninept
This paper addresses the problem of blind stereoscopic image quality assessment (NR-SIQA) using a new multi-task deep learning based-method. In the field of stereoscopic vision, the information is fairly distributed between the left and right views as well as the binocular phenomenon. In this work, we propose to integrate these characteristics to estimate the quality of stereoscopic images without reference through a convolutional neural network. Our method is based on two main tasks: the first task predicts naturalness analysis based features adapted to stereo images, while the second task predicts the quality of such images. The former, so-called auxiliary task, aims to find more robust and relevant features to improve the quality prediction. To do this, we compute naturalness-based features using a Natural Scene Statistics (NSS) model in the complex wavelet domain. It allows to capture the statistical dependency between pairs of the stereoscopic images. Experiments are conducted on the well known LIVE PHASE I and LIVE PHASE II databases. The results obtained show the relevance of our method when comparing with those of the state-of-the-art. Our code is  available online on \url{https://github.com/Bourbia-Salima/multitask-cnn-nrsiqa_2021}.
\end{abstract}
\vspace{-4mm}
\section{Introduction}
\label{sec:intro}
\vspace{-4mm}
Nowadays, stereoscopic images (SI) are commonplace in a wide range of applications such as 3D movies, medical imaging, virtual reality, 3D video games and so forth. Usually, several operations are applied to SI (restoration, compression, transmission, etc.), each can affect the perceived quality of such images. This issue has motivated the computer vision community to propose sophisticated quality metrics that predict the perceptual impact of these distortions, so-called Stereo Image Quality Assessment (SIQA). Two approaches are adopted: subjective and objective \cite{mohammadi2014subjective}. Subjective quality evaluations are based on the human perception. Apparently, this kind of evaluation is time-consuming and consequently unpractical in real-world applications. On the other hand, the objective quality evaluations aim to predict automatically the quality scores with computational models. Depending on the presence of reference SIs, there are three types of objective measures: Full Reference SIQA (FR-SIQA) that requires the presence of the reference image to evaluate the quality, Reduced Reference SIQA (RR-SIQA) that requests reduced information of the original image, and No Reference (NR-SIQA) that assesses the quality without the need of any information from the reference image. Thanks to the requirement of most nowadays applications and with the considerable development of machine learning techniques, most of the existing SIQA methods are dedicated to NR-SIQA. The first NR-SIQA methods extracted several handcrafted features from the distorted SI. Then fed them into a regression method (logistic function, support vector regression,etc.) \cite{appina2016no,su2015oriented,shao2015blind, fezza2017using,VCIP14Chetouani}. Recently, motivated by the promising results of deep learning in different image processing and computer vision fields \cite{VCIP14,SPIC20Chetouani,ICIP18Ilyass, PRL20CHETOUANI,info19Hamidi,MTAP19Fourati} and so on), this approach was naturally extended to SIQA. Zhang et al. \cite{zhang2016learning} used the left and right images as well as the corresponding difference image as an input of a Convolutional Neural Network (CNN). The convolutions and the max-pooling applied to these three images are concatenated and used by a multi layer perceptron network to estimate the quality score. In the work of Chetouani et al. \cite{chetouani2018blind}, the authors proposed a two-step framework. The type of the degradation is identified in the first step using a CNN model while the quality score is computed in the second step by fusing features according to the identified degradation type. This fusion is achieved by using a support vector regression model. Based on a CNN, Zhou et al. \cite{zhou2019dual} proposed a dual-stream network composed by two sub-networks. Each one corresponds to the left and right view, respectively. Additionally, the authors record the interaction of these two sub-networks in multiple layers to take into account the binocular information. However, this method remains insufficient since the statistical dependency between the two views was not considered.

To overcome this limitation, we propose in this study a multi-task CNN for NR-SIQA. The idea is to extract  features based on naturalness analysis that model the statistical dependence of the stereo pair. The latter aims to help the CNN to better predict the perceptual quality of stereo images. These features are integrated in the proposed scheme as an auxiliary task.

The remainder of this paper is organized as follows. In Section \ref{sec2}, we describe the proposed method. Then, we present the experimental results in the Section \ref{sec3}. Finally, we give some concluding remarks in Section \ref{sec4}.

\vspace{-5mm}
\section{Proposed deep-based blind stereo image quality assessment method}
\label{sec2}
\begin{figure*}[h]
\centering
\includegraphics[scale=0.8]{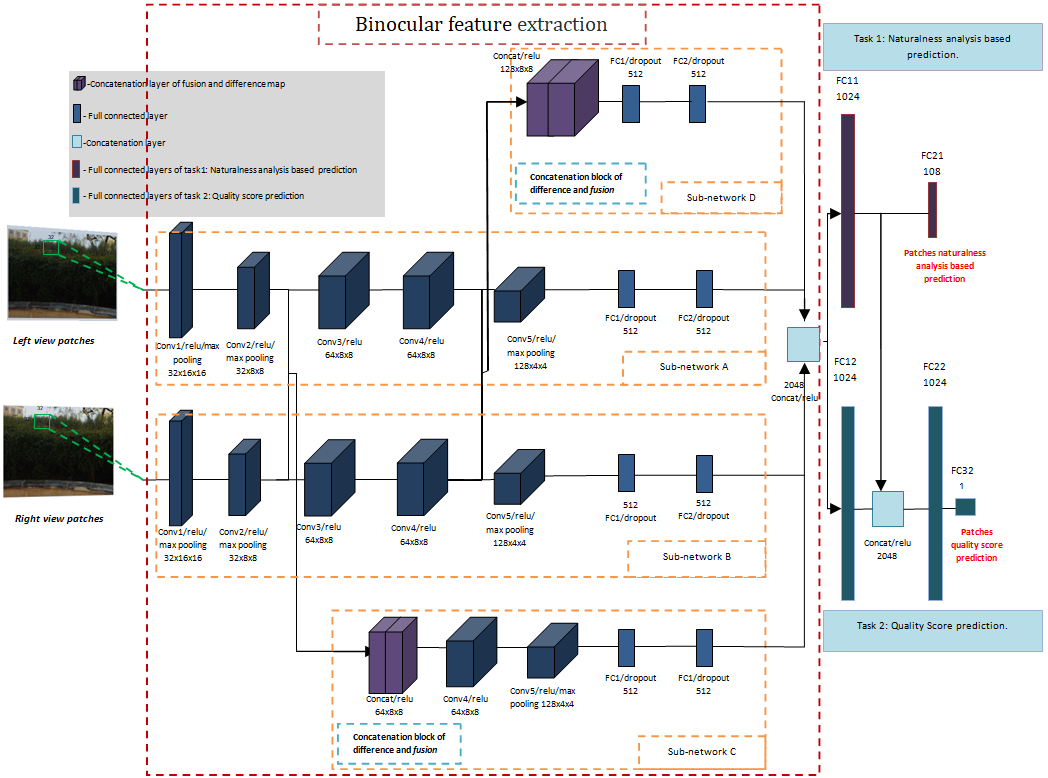}
\vspace{-4mm}
\caption{Flowchart of the proposed NR-SIQA method.}
\label{t}
\end{figure*}

\vspace{-3mm}
Fig. \ref{t} presents the flowchart of the proposed method which has two major stages: Binocular feature extraction and multi-task prediction. The first step aims to extract features that characterize each view of SI as well as their binocular fusion and disparity responses. Whereas, the second step computes the final quality score of SI through a multi-task prediction. These steps are described in this section. 
\vspace{-3mm}
\subsection{Deep Learning-based Feature extraction}
\label{sec:arch}
\vspace{-2mm}
In this step, we adopt a CNN architecture which requires a large number of images to be trained in order to avoid the problem of over-fitting. This requirement can not achieved using the limited SI datasets often used for the quality assessment. So, we trained our network with normalized patches of size $32\times32$ with a stride of $24$, extracted from distorted stereoscopic images. Each patch has the same score as its corresponding source in the stereoscopic image. Inspired by the work of \cite{zhou2019dual}, features are extracted from four sub-networks as shown in Fig. \ref{t}. The sub-networks A and B extract features from the left and the right distorted images, respectively by using five convolutional layers and two fully connected layers. For the sub-network C, we concatenate the results of the two first convolutional layers of the sub-networks A and B in low CNN layers and we add two convolutional layers and two fully connected layers. Similarly, in the sub-network D, the results of the fourth convolutional layers of the sub-networks A and B in a high CNN layers are concatenated and used as an input of two fully connected layers without convolution operation. The sub-networks C and D represent the disparity and the feature fusion map. Finally, the vectors extracted from the sub-networks are concatenated into one vector of size $2048$ that allows to model the binocular characteristics of the Human Visual System (HVS). 

\begin{table*}
\caption{Comparison study of our method with state-of-the-art methods on LIVE PHASE I database. The first two higher performances are highlighted in bold.} 
\resizebox{0.8\textwidth}{!}{\begin{minipage}{\textwidth}
\begin{center}
\begin{tabular}{|*{ 14}{c|}} 
    \hline 
\multirow{2}{*}{Type}  & \multirow{2}{*}{Methods}  & \multicolumn{2}{|c|}{JP2K } &  \multicolumn{2}{|c|}{JPEG } &  \multicolumn{2}{|c|}{WN} &   \multicolumn{2}{|c|}{BLUR } & \multicolumn{2}{|c|}{FF }  &\multicolumn{2}{|c|}{ALL }   \\ 
   
  &   &    PLCC   & SROCC &    PLCC   & SROCC   & PLCC   & SROCC & PLCC & SROCC & PLCC & SROCC &    PLCC & SROCC    \\

    \hline         
      \multirow{2}{*}{FR}  &    Benoit et al. \cite{4711773}  &  0.939  & \textbf{0.910} &  0.640 &  0.602 & 0.925  & 0.929  &   0.948 & \textbf{0.930}  &  0.747  &  0.698 & 0.902  &   0.889   \\  

  & You et al.\cite{inproceedings}  & 0.877  & 0.859  & 0.487 & 0.438 &  0.941 & 0.939 & 0.919 & 0.882 & 0.730 & 0.583& 0.881 &  0.878\\
    \hline  
    \hline 
      \multirow{2}{*}{RR}   & Wang et al.\cite{wang2015image}  & 0.916  &  0.883& 0.569 & 0.542  & 0.913 & 0.906 & 0.957  & \textbf{0.924}  & 0.783 & 0.654 & 0.892 & 0.889  \\ 
    & Ma et al.  \cite{ma2016reorganized}  & 0.918 & 0.886 & 0.722  &0.616 &0.913 & 0.912 &  0.924 & 0.879 & 0.806  &0.696  &0.905 & 0.905 \\ 

    \hline 
    \hline 
       \multirow{6}{*}{NR}  &   Akhter et al. \cite{akhter2010no}   & 0.905  & 0.866 & 0.729 & 0.675 &  0.904 & 0.914 &  0.617 & 0.555 &  0.503 & 0.640  &  0.626 & 0.383 \\ 

    & Shao et al. \cite{shao2015blind}   & 0.872  & 0.900  &   \textbf{0.897 }& 0.607  &   0.916 &  0.903  &  0.923 & 0.923  &  —–   &  —–   &  0.899 &0.894     \\ 
   & Chen et al.\cite{chen2013no} & 0.907  &  0.863 & 0.695  & 0.617  &  0.917 & 0.919  &  0.917  & 0.878  &   0.735 & 0.652  &  0.895  &  0.891 \\ 

    & Zhang et al.\cite{zhang2016learning} & 0.926   &\textbf{ 0.931}   & 0.740  & \textbf{ 0.693} & 0.944  & \textbf{ 0.946} & 0.930  &  0.909  &  \textbf{0.883}  &   \textbf{0.834} &0.947   &\textbf{ 0.943 } \\

   & Zhou et al. \cite{zhou2019dual}  &\textbf{ 0.955}  &  0.896 &  0.714 &  0.650 &\textbf{ 0.960}  &  0.945 &\textbf{ 0.956}  &  0.836 &\textbf{ 0.849}  & \textbf{ 0.811 } &\textbf{0.953 }& 0.935   \\  

    & \textbf{The proposed NR-SIQA}  & \textbf{ 0.951} & 0.908  &  \textbf{0.744 }& \textbf{ 0.679}  & \textbf{0.966  } &  \textbf{0.949}  & \textbf{ 0.972}  &  0.867  &   0.838 & 0.782 & \textbf{ 0.957}  &\textbf{ 0.942 }  \\ 

    \hline 

\end{tabular}
\end{center}  
\end{minipage}}
\label{phase1} 
\end{table*}

\begin{table*}
\vspace{-7mm}
\caption{Comparison study of our method with state-of-the-art methods on LIVE PHASE II database. The first two higher performances are highlighted in bold.} 
\resizebox{0.8\textwidth}{!}{\begin{minipage}{\textwidth}
\begin{center}
\begin{tabular}{|*{ 14}{c|}} 
 \hline  
\multirow{2}{*}{Type}   & \multicolumn{1}{|c|}{\multirow{2}{*}{Methods}}  & \multicolumn{2}{|c|}{JP2K } &  \multicolumn{2}{|c|}{JPEG } &  \multicolumn{2}{|c|}{WN} &   \multicolumn{2}{|c|}{BLUR } & \multicolumn{2}{|c|}{FF }  &\multicolumn{2}{|c|}{ALL }   \\ 
      &    &    PLCC   & SROCC &    PLCC   & SROCC   & PLCC   & SROCC & PLCC & SROCC & PLCC & SROCC  & PLCC & SROCC       \\ 

    \hline 
     \multirow{2}{*}{FR}   &  Benoit et al. \cite{4711773}  & 0.784  &  0.751 &  0.853 & \textbf{0.867}  & \textbf{0.926 } & \textbf{0.923 } & 0.535   & 0.455  &   0.807  &  0.773  &  0.748  & 0.728  \\ 
     
       &  You et al. \cite{inproceedings} &  \textbf{0.905} & 0.894  &  0.830 & 0.795  & 0.912  & 0.909  &  0.784 &  0.813  &  0.915 &  0.891 &  0.800 &  0.786  \\  
    \hline
    \hline

    \multirow{4}{*}{NR}   &    Akhter et al. \cite{akhter2010no}  & 0.776  &  0.724 &  0.786 & 0.649  &  0.722 & 0.714  &   0.795  &   0.682   &  0.674   &  0.559  &  0.568 &  0.543 \\ 

        &  Chen et al. \cite{chen2013no}  & 0.899  &  0.867 & \textbf{0.901}  &\textbf{ 0.867}  & \textbf{0.947}  &\textbf{ 0.950}  &  0.941  &  0.900 &\textbf{ 0.932 } &  \textbf{0.933} &   0.895 &   0.880 \\ 
 
       &  Zhou et al. \cite{zhou2019dual}  & 0.889  & \textbf{0.902}  &  0.872 &  0.860 & 0.884  & 0.874& \textbf{ 0.978} &   \textbf{0.908 }& 0.929  &\textbf{ 0.920} &\textbf{ 0.902 }&  \textbf{0.897} \\  

        &  \textbf{The proposed NR-SIQA} &  \textbf{0.912} &  \textbf{0.915} & \textbf{ 0.874} &  \textbf{0.874} & 0.924& 0.889  &\textbf{ 0.986 } &\textbf{ 0.924}  & \textbf{0.948}  & \textbf{ 0.933} & \textbf{ 0.921 }& \textbf{0.915 } \\  

    \hline 
\end{tabular} 
\end{center}  
\end{minipage}}
\label{phase2} 
\end{table*} 

\vspace{-4mm}
\subsection{Multi-task prediction}
\vspace{-3mm}
The vector extracted is then fed as an input to two tasks: a main task (Task2) that predicts the quality score and an auxiliary task (Task1) that predicts the naturalness analysis based-features. The Task1 is mainly used to help the Task2 to improve the quality prediction.

\vspace{-5mm}
\subsubsection{Task 1: Naturalness based-features prediction}
\vspace{-3mm}
The SI has certain regular statistical properties following a certain type of distributions, that can be impacted with the presence of a distortion. 
Additionally, there is a high dependence between the left and right images of the stereo pair as demonstrated in the work of \cite{karine2018novel}.
In this study, these specificities are considered by exploiting of a multivariate statistical model. We measure the naturalness degree of stereo images through NSS based features. Besides, we capture the dependency between the left and right views.
More precisely, we use a copula statistical model by first applying the dual-tree complex wavelet transform (DT-CWT) method \cite{DBLP:journals/lgrs/KarineTKH17} to each view of the distorted stereo pair, since the DT-CWT subband decomposition is a combination of band-pass filters, and the filter response mimics the space orientation in the V1 area of primary visual cortex. 
After that, we model the magnitude of the resulting complex wavelet coefficients using the copula. The main advantage of a copula-based model is that captures the dependence between wavelet coefficients while maintaining a good fit to the marginal distributions. In our work, we use a Gaussian copula : 
\begin{equation}
    f(\overrightarrow{x},\theta ) = \frac{1}{\lvert \sum ^{1/2} \lvert} \exp{ \frac{-\overrightarrow{y}^{t} ( \sum ^{-1} - I) \overrightarrow{y}}{2} } \times \prod_{i=1}^d f_{i}(x_{i} ; p)
\label{copulas}
\end{equation}
with the  Gamma distribution as a margin : 
\begin{equation}
    f(x,p = (a , b)) = \frac{a^{-b} x^{b-1}}{\Gamma{(b)}} \exp{- \left ( \frac{x}{a}\right ) }  , x \in {R}^{+}
\label{Gamma}
\end{equation}
where $\overrightarrow{x}  =  (x_{1},...,x_{d})$  is a vector of $d$ randomly and independently selected DT-CWT coefficients,  $\theta  = (p ,\sum)$ denotes the copula hyper-parameters, $p= ( b , a) $ is the margins parameters vector while $b>0$ is the shape and $a>0$ is the scale parameter, $\sum$ indicates the covariance matrix,  $\overrightarrow{y}$ represent the Gaussian vector given by $y_{i} = \phi^{-1} (F(x_{i}; p_{1},p_{2} )$  ( $\phi$ is the cumulative distribution function (CDF) of normal distribution $N (0,1))$), $d$ is the dimension of copula and $f_{i}$  is the margin defined in our method by Eq.\ref{Gamma} that demonstrate the gamma marginal distribution. The univariate and multivariate statistical parameters are estimated through the Maximum-Likelihood algorithm. More details can be found in \cite{karine2018novel,8518668, article}.

Once the multivariate statistical parameters are computed, they will be used as labels for each input stereo pair. This part is composed of two fully connected layers, FC11 and FC21 of size $1024$ and $108$, respectively. The last layer (i.e. FC21) of this task predicts the naturalness based-features of each patch.

\vspace{-5mm}
\subsubsection{Task 2: Quality score prediction} 
\vspace{-3mm}
This task is composed of three fully connected layers (i.e. FC12, FC22 and FC32). FC12 is of size $1024$ and has as input the above-extracted deep learning-based features (see Section \ref{sec:arch}). With the same size, FC22 has as input the concatenation of the feature vectors generated by the layers FC12 and FC11. This concatenation aims to get more significant features to better predict the quality score through the features given by FC11. The last fully connected layer, FC32, is of size $1$ and offers the predicted Differential Mean Opinion Score ($DMOS$) of each extracted patch. The quality score of the whole image is finally given by computing the mean predicted patch scores.

\vspace{-4mm}
\subsection{Training}
\vspace{-2mm}
To minimize the error during the training of the designed CNN model, we use the loss function $\mathcal Loss$ which is a linear combination of two $L_1$ loss functions as described in Eq.\ref{loss}:
\begin{equation}
    \mathcal Loss = \lambda \lvert \hat{Y}_2 - Y_2 \rvert + \lvert \hat{Y}_{1} - Y_1 \rvert
\label{loss}
\end{equation}
where $Y_{2}$ and $\hat{Y}_2$ are the subjective quality ($DMOS$) and the predicted quality scores of each patch, respectively used for Task 2. $\hat{Y}_{1}$ is the predicted naturalness analysis based-features and $Y_{1}$ is the corresponding ground truth naturalness analysis based-features used for Task 1. $\lambda$ represents an adjustment coefficient that balances the losses between the two tasks. This parameter is tuned experimentally and it was set to 25.

To update the weights of the whole network, we used the Stochastic Gradient Descent (SGD) with a momentum factor equals to 0.9, a weight decay factor sets to 0.0001, a mini batch size equals to $128$ and a learning rate initialized to $10^{-3}$. Our method was implemented using Pytorch framework.  
\vspace{-2mm}
\section{Experimental results}
\label{sec3}
\vspace{-3mm}
\subsection{Stereo image quality databases}
\label{database}
\vspace{-3mm}
The performance of our method was quantified using two databases. The first one is LIVE PHASE I \cite{moorthy2013subjective} which is composed of $20$ reference SIs and their corresponding $365$ symmetrically distorted version obtained with five various distortions. The second database, LIVE PHASE II \cite{chen2013full}, contains $360$ distorted SI derived from $8$ reference SIs. The distortions are obtained by symmetrically and asymmetrically applying five different distortions. $DMOS$ values are provided for each distorted SI of both databases where low quality value corresponds to a high value of $DMOS$. 

To evaluate the capacity of our method to predict the quality of such images, we applied the protocol evaluation described in \cite{zhang2016learning} and all the methods that we used in the performance comparison (Section.\ref{sec:compar}) followed the same protocol (the method of Zhou et al. \cite{zhou2019dual} was  retrained on the same protocol). More precisely, we conduct a hold-out cross validation by dividing 10 times each SI database with 60\% of the SI used to train our model, 20\% for validation and 20\% for test. In order to ensure that our model assesses the quality of the image and not of the content, the SI used in the training phase are independent from those used in the validation and test phases.

\vspace{-7mm}
\subsection{Performance comparison}
\label{sec:compar}
\vspace{-3mm}
The efficiency of the proposed method was demonstrated through two criteria: Pearson Linear Correlation Coefficient (PLCC) to measure the prediction accuracy and Spearman Rank Order Coefficient (SROCC) to measure the prediction monotonicity. For both criteria, a higher absolute value indicates a better prediction performance. We compare in Tables.\ref{phase1} and \ref{phase2} the performance of our proposed method with different state-of-the-art methods using the average of the PLCC and SROCC of LIVE PHASE I and LIVE PHASE II, respectively. Many observations can be raised from these tables. First, generally NR-SIQA methods outperform FR-SIQA and RR-SIQA. Second, the deep learning-based approaches \cite{zhang2016learning,zhou2019dual} show the highest correlation coefficients comparing to the other learning-based methods \cite{akhter2010no,shao2015blind,chen2013no}. Third, our method achieves a competitive performance for the majority of the distortion types. Finally, the proposed method presents the higher overall PLCC on the two databases. Furthermore, we investigate in Table.\ref{phase2_asy} the performance against the fairness of distortion types (symmetric or asymmetric) for the LIVE PHASE II database. For both, our method achieves the best SROCC. These results demonstrate the robustness of the proposed method against the degradation discrepancy for SI-QA. In order to test the effect of the auxiliary task (i.e. naturalness based-features prediction) on the main task of our method (i.e. quality score prediction), we conduct in Table.\ref{ablation} an ablation study by cancelling auxiliary task including the concatenation part. The results demonstrate that the performances are improved with the addition of the auxiliary task. The explanation for this improvement is that the statistical dependence between SI helps the CNN in the training phase. To verify the accuracy prediction of the Task 1, we calculated the Root Mean square Error (RMSE) between the real and the predicted naturalness analysis based-features. It equals to $0.391$ for LIVE PHASE I and $0.416$ for LIVE PHASE II. 
\vspace{-4mm}

\section{Conclusion}
\label{sec4}
\vspace{-4mm}
In this paper, we presented a new method based on multi-task convolutional neural network model to evaluate the quality of NR-SIQA. We used naturalness analysis based-features prediction as an auxiliary task to improve the quality prediction. Based on the comparative study on two public databases (i.e. LIVE PHASE I and LIVE PHASE II), our model is competitive to state-of-the-art methods, especially in the case of asymmetrical distortions. The ablation study shows that adopting an auxiliary task based on naturalness analysis based-features represents a promising track for an effective enhancement of the prediction of quality scores. Further work will concern the extension of the model by using more perceptual attributes as auxiliary task. 

\begin{table}
\vspace{-8mm}
\caption{Performance study of our method against the distortion distribution on LIVE PHASE II database (SROCC).}
\vspace{-7mm}
\begin{center}
\resizebox{0.85\textwidth}{!}{\begin{minipage}{\textwidth}
\begin{tabular}{|*{ 4}{c|}} 
 \hline 
     Type     & Methods     &    Symmetric   & Asymmetric   \\ 

    \hline 
   \multirow{2}{*}{FR}  &   Benoit et al. \cite{4711773} &  0.860 & 0.671 \\  
    
     &   You et al. \cite{inproceedings} &0.914 &  0.701\\  
  \hline
  \hline
  
    \multirow{4}{*}{NR}    &    Akhter and al. \cite{akhter2010no} &  0.420  &0.517 \\ 

    &   Chen et al. \cite{chen2013no} & \textbf{ 0.918}  & 0.834  \\

    &  Zhou et al. \cite{zhou2019dual}  &  0.911 &  0.869  \\  

    & \textbf{ The proposed NR-SIQA}  & \textbf{0.918}  &  \textbf{ 0.896}    \\  
    \hline 
\end{tabular} 
\end{minipage}}
\end{center} 
\label{phase2_asy} 
\end{table}

\begin{table}
\vspace{-9mm}
\caption{Ablation study on LIVE PHASE I and PHASE II.}
\vspace{-3mm}
\begin{center}
\resizebox{0.95\textwidth}{!}{\begin{minipage}{\textwidth}
\begin{tabular}{|*{8}{c|}}
    \hline
          &   \multicolumn{2}{|c|}{LIVE PHASE I }  &  \multicolumn{2}{|c|}{LIVE PHASE II }\\
           
                  & PLCC  & SROCC    & PLCC  & SROCC    \\
                                                 
    \hline
     Model without & & & & \\  the auxiliary task  &0.953 & 0.935  & 0.902  &  0.897     \\  

    \hline
   Full model &  \textbf{0.957} & \textbf{ 0.942} & \textbf{ 0.921} & \textbf{0.915} \\
    \hline
\end{tabular}
\end{minipage}}
\end{center}
\label{ablation}
\end{table}
\vspace{-5mm}

{\small
\bibliographystyle{IEEEbib}

}
\end{document}